\documentclass[francais]{gretsi}

\usepackage[utf8]{inputenc}

\usepackage[T1]{fontenc}

\usepackage{amsmath, amssymb, amsfonts}
\usepackage{bm}

\usepackage{xcolor}
\usepackage{siunitx}
\usepackage{multirow}
\usepackage[normalem]{ulem} 

\def\rtsixty{\text{RT}_{60}}

\def\C{\mathbb{C}}

\def\thineq{\hspace{-.1em}=\hspace{-.1em}}

\begin{document}


\titre{
Déréverbération non-supervisée de la parole par modèle hybride
}

\auteurs{
  \auteur{Louis}{Bahrman}{louis.bahrman@telecom-paris.fr}{}
  \auteur{Mathieu}{Fontaine}{mathieu.fontaine@telecom-paris.fr}{}
  \auteur{Gaël}{Richard}{gael.richard@telecom-paris.fr}{}
}

\affils{
  \affil{}{LTCI, T\'el\'ecom Paris, Institut Polytechnique de Paris, Palaiseau, France
  }
}

\resume{
Cet article introduit une nouvelle stratégie d'apprentissage pour améliorer des systèmes de déréverbération de la parole de manière non-supervisée en n'utilisant que des signaux réverbérants. 
La plupart des algorithmes existants nécessitent des paires de signaux (sec, réverbérant), qui sont difficiles à obtenir.
Notre approche utilise en revanche des informations acoustiques limitées, comme le temps de réverbération (RT60), pour entraîner un système de déréverbération.
Les résultats expérimentaux démontrent que notre méthode permet d'obtenir des performances plus cohérentes que l'état de l'art sur différentes mesures objectives.
}

\abstract{
This paper introduces a new training strategy to improve speech dereverberation systems in an unsupervised manner using only reverberant speech. 
Most existing algorithms rely on paired dry/reverberant data, which is difficult to obtain.
Our approach uses limited acoustic information, like the reverberation time (RT60), to train a dereverberation system.
Experimental results demonstrate that our method achieves more consistent performance across
various objective metrics than the state-of-the-art.
}

\maketitle

\section{Introduction}
Les signaux acoustiques capturés dans des salles sont affectés par des réflexions par les murs et la diffraction par des obstacles rencontrés sur le chemin acoustique, dans un processus dénommé réverbération, qui réduit l'intelligibilité des enregistrements de parole,
et justifie la nécessité d'employer des méthodes de déréverbération pour les atténuer.
La tâche de déréverbération a été historiquement résolue en utiliant des méthodes statistiques de traitement du signal~\cite{nakatani_speech_2010}.
L'absence de solution unique au problème de déréverbération encourage l'usage de réseaux neuronaux profonds (RNP), qui requièrent en pratique de grandes quantités de données.

Ces approches peuvent être supervisées de différentes manières. 
Les approches discriminatives apprennent à prédire un signal sec~\cite{wang_tf-gridnet_2023}, ou un masque complexe~\cite{hao_fullsubnet_2021} à partir d'un signal réverbérant, et requièrent une grande quantité de données par paires (sèches,réverbérantes). 
Les modèles génératifs, comme les auto-encodeurs variationels~\cite{leglaiveRecurrentVariationalAutoencoder2020} 
apprennent la distribution de signaux secs sans avoir accès à des signaux réverbérants durant l'entraînement.
Bien que ces modèles nécessitent moins de supervision, ils ne résolvent pas le problème d'accès aux données, car les données sèches sont plus difficiles à obtenir que les données réverbérantes.
Ainsi, des approches exploitant uniquement des signaux réverbérants ont été conçues, dont
MetricGAN-U~\cite{fuMetricGANUUnsupervisedSpeech2022}. Son paradigme d'entraînement est basé sur un réseau antagoniste (GAN) dont le discriminateur est entraîné à imiter une métrique cible, et le générateur à optimiser sa performance vis-à-vis du discriminateur.
Cette approche a été appliquée avec succès à la déreverbération en utilisant la métrique du \textit{rapport parole à énergie réverbérante} (SRMR)~\cite{falkNonIntrusiveQualityIntelligibility2010} en tant que métrique cible à optimiser.

\begin{figure}[t]
    \centering
    \includegraphics[width=.9\linewidth]{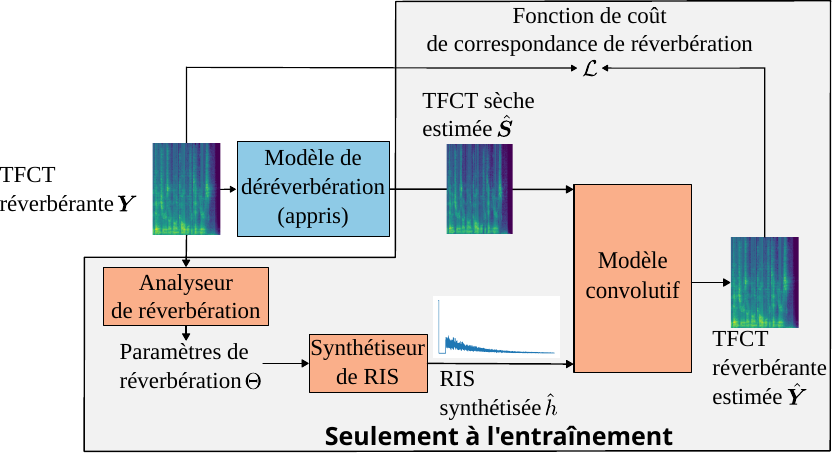}
    \vspace{-0.5em}
    \caption{Aperçu de la méthode proposée}
    \label{fig:overview}
    \vspace{-1em}
\end{figure}

De plus, les approches supervisées et non supervisées pour la déréverbération ont été améliorées en les hybridant avec des modèles de réverbération classiques.
Un choix populaire pour modéliser implicitement la réverbération est l'approximation de la fonction de transfert convolutive (CTF), qui considère la réverbération comme un processus de filtrage en sous-bandes. Elle a été utilisée dans la méthode de l'erreur de prédiction pondérée (WPE)~\cite{nakatani_speech_2010}.
Un modèle établi à partir de la CTF
a même été utilisé pour la déréverbération supervisée uniquement par le signal réverbérant dans USDNet~\cite{wangUSDnetUnsupervisedSpeech2024a}.
L'énergie dans chacune des bandes peut être modélisée par une décroissance exponentielle, et, 
dans~\cite{molinerBUDDySingleChannelBlind2024}, les paramètres de cette décroissance et le signal sec sont alternativement estimés par un modèle de diffusion.
Certains modèles ont même été conçus pour déréverbérer en ayant accès à l'inférence aux propriétés de la réverbération~\cite{wu_reverberation-time-aware_2017}.
L'essor de ces méthodes a été permis par des avancées significatives en estimation aveugle, c.-à-d. à partir du signal réverbérant, des paramètres de réverbération. 
Le temps de réverbération, qui décrit la décroissance de l'énergie de la RIS, peut notamment être estimé grâce à un algorithme fondé sur la décomposition en sous-bandes du spectrogramme du signal réverbérant~\cite{prego_2015_blind}.
Jusqu'à présent, MetricGAN-U était la meilleure approche de déréverbération supervisée par les signaux réverbérants, surpassant WPE et USDNet.
Nous qualifions cette approche d'\textit{auto-supervision par une métrique}.

Dans cet article, nous proposons d'introduire un nouveau cadre hybride pour la déréverbération non-supervisée, appelé \textit{auto-supervision par réverbération}.
Nous entraînons un RNP à estimer un signal de parole sèche, de telle sorte qu'un modèle de réverbération appliqué sur ce signal estimé corresponde à son signal d'entrée réverbérant.
Nous montrons, pour diverses mesures objectives, que la déréverbération auto-supervisée par réverbération est plus performante que la déreverbération basée sur les métriques. 
À des fins de reproductibilité et pour faciliter les recherches futures, nous distribuons publiquement des exemples, le code et les modèles pré-entraînés\footnote{\url{https://louis-bahrman.github.io/Hybrid-WSSD/}}.

\section{Modèle de réverbération}

\subsection{Réverbération tardive}
En supposant des positions de source et de microphone fixes, un signal monaural réverbérant $y$ peut être représenté comme la convolution d'un signal sec $s$ et une réponse impulsionelle de salle (RIS) entre la source et le microphone $h$:
\begin{equation}
    y(n) = (s \star h)(n), \label{eq:time_domain_convolution}
\end{equation}
où $n$ dénote l'index temporel et $\star$ l'opérateur de convolution.
La RIS $h$ peut être divisée en 3 parties: le trajet direct correspond à son premier pic $h_d$ suivi des réflexions précoces $h_e$ et, après le temps de mixage $n_m$, la réverbération tardive $h_l$.

Un modèle simple de réverbération tardive est le modèle de Polack~\cite{polackTransmissionEnergieSonore1988}.
Ce modèle considère $h_l$ comme la réalisation d'un bruit blanc sous enveloppe exponentiellement décroissante:
\begin{equation}
h_l(n) = b(n) e^{\frac{-3 \ln(10) n }{\rtsixty f_s}}, 
\end{equation}
avec $b(n) \sim \mathcal{N}(0, \sigma^2)$ une distribution normale centrée, $\rtsixty$ le temps de réverbération et $f_s$ la fréquence d'échantillonage.

\subsection{Convolution dans le plan temps-fréquence}
Le système invariant de l'Eq. \eqref{eq:time_domain_convolution} peut être formulé comme une convolution inter-bande et inter-trame dans le domaine de la Transformée de Fourier Court-Terme TFCT~\cite{avargel_system_2007}:
\begin{equation}
Y_{f,t}=\sum_{f^{\prime}=0}^{F-1}\sum_{t^{\prime}=0}^{\min(t; T_h)} \mathcal{H}_{f,f^{\prime},t^{\prime}} S_{f^{\prime},t-t^{\prime}}, \label{eq:full_convolution}
\end{equation}
où $\bm{Y} \triangleq \{Y_{f,t}\}_{f,t=0}^{F-1, T_y-1} \in \mathbb{C}^{F \times T_y}$ sont les coefficients de la TFCT du signal réverbérant à la fréquence $f\thineq 0,\dots,F-1$ et à la trame $t\thineq 0,\dots,T_y-1$, 
$\mathcal{H} \triangleq \{\mathcal{H}_{f, f^\prime, t}\}_{f,f^\prime, t=0}^{F-1,F-1,T_h-1} \in \C^{F \times F \times T_h}$ est une représentation tridimensionnelle de la RIS
et $\bm{S} \triangleq \{S_{f,t}\}_{f,t=0}^{F-1, T_s-1}\in \C^{F\times T_s}$ est la TFCT du signal sec.
Comme démontré dans~\cite{avargel_system_2007}, $\mathcal{H}$ peut être calculé à partir de $h \in \mathbb{R}^{N_h}$ comme : 
\begin{equation}
    \mathcal{H}_{f, f', t'}
    =
    \sum_{m=-N+1}^{N-1} h(t'L - m) W_{f,f^\prime}(m), \label{eq:big_H}
\end{equation}
où $N$ est la longueur de fenêtre de TFCT, $L$ la taille de saut et
\begin{equation}
     W_{f,f^\prime}(m)=\frac{1}{F}\sum_{n=0}^{N-1}
    w_s(n + m)w_a(n) e^{\frac{j2\pi (f^\prime (n+m) - fn)}{F}} \label{eq:big_window} 
\end{equation}
 avec $w_s, w_a$ les fenêtres d'analyse et de synthèse respectives.

\section{Méthode}

\subsection{Aperçu}

Nous proposons d'entraîner un modèle d'apprentissage profond de déréverbération en le supervisant par un modèle de réverbération.
La procédure d'entraînement est la suivante:
Étant donné un signal réverbérant $\bm{Y}$ défini à la section précédente, le RNP renvoie un signal sec estimé $\hat{\bm{S}} \triangleq \{\hat{S}_{f,t}\}_{f,t=0}^{F-1, T_s-1}\in \C^{F\times T_s}$.
En parallèle, un modèle de réverbération $\mathcal{R}$, estime à partir du signal réverbérant le temps de réverbération $\rtsixty$ et s'en sert pour synthétiser une RIS approximée $\hat{h} \in \mathbb{R}^{N_h}$.
La TFCT du signal sec estimé $\hat{\bm{S}}$ est ensuite convoluée avec la RIS synthétique $\hat{h}$ grâce au modèle inter-bande $\mathcal{C}$ (cf. Eq.~\eqref{eq:cb_convol_model}), pour estimer la TFCT du signal réverbérant $\hat{\bm{Y}}$.
La fonction de coût de déréverbération nécessitant des paires de signaux secs et réverbérants est remplacée par une fonction de coût de correspondance de réverbération $\mathcal{L}$, qui calcule la distance entre le spectrogramme estimé $\hat{\bm{Y}}$  et la référence $\bm{Y}$.
Un schéma de la procédure est présenté à la Fig.~\ref{fig:overview}.
Étant donné que le modèle de synthèse de RIS et le modèle convolutif ne sont pas paramétriques, ils n'ont pas besoin d'être entraînés. 
À l'inférence, seul le RNP est utilisé, et ainsi le nombre de paramètres ainsi que la complexité temps et mémoire demeurent les mêmes que pour le RNP de déréverbération.

\subsection{Modèle de RIS}

Le modèle de RIS sert à synthéthiser une RIS dont les caractéristiques sont celles de la RIS correspondant à la vérité terrain.
Il se décompose en 2 parties, une d'analyse dénotée $\mathcal{A}$, visant à estimer les paramètres acoustiques à partir du signal réverbérant, et une de synthèse, dénotée $\mathcal{S}$, visant à synthétiser une RIS réaliste à partir de ces caractéristiques.

Le synthétiseur de RIS sert à synthéthiser une RIS dont la réverbération tardive $h_l$ correspond au modèle de Polack et le trajet direct $h_d$ est un pic d'amplitude $1$. 
Pour mieux faire correspondre le modèle à la distribution de nos données sans modifier la distribution de l'énergie décrite par Polack,
et suite à des expériences préliminaires, nous avons décidé de synthétiser une RIS en utilisant la valeur absolue de la distribution gaussienne utilisée dans le modèle de Polack.
Afin d'aligner les signaux secs et réverbérants, nous supprimons les échantillons de RIS précédent le premier pic.
Ainsi, la RIS synthéthique devient:
\begin{equation}
    \mathcal{S}(\Theta)(n) = \begin{cases}
        | b(n) | e^{-\frac{3 \ln(10)}{\rtsixty f_s} n } & \text{si } n > n_m\\
        1 & \text{si } n = 0\\
        0 & \text{sinon},   
    \end{cases}
\end{equation}
où $b(n)$ est tiré d'une distribution normale $\mathcal{N}(0, \sigma^2)$.
Durant l'entraînement, une RIS est synthétisée à partir d'un nouveau tirage de bruit à chaque pas de gradient.

\subsection{Modèle convolutif et fonction de coût}

Afin de mieux rétropropager le gradient au modèle de déréverbération dont la sortie peut être dans le plan temps-fréquence, nous considérons un modèle convolutif inter-bande en temps-fréquence et une fonction de coût de correspondance de réverbération. 
Étant donné $\hat{h}=\mathcal{S}(\Theta)$ et $\hat{\bm{S}}$ le signal sec estimé par le RNP, nous définissons le modèle de convolution temps-fréquence comme:

\begin{align}
    \hat{\bm{Y}}_{f,t} \triangleq \mathcal{C}(\hat{\bm{S}}, \hat{h}) 
    = 
    \sum_{f^{\prime}=f-F^{\prime}}^{f+F^{\prime}}\sum_{t^{\prime}=0}^{\min(t; T_h)} \hat{\mathcal{H}}_{f, f', t'}\hat{S}_{f^{\prime},t-t^{\prime}} \label{eq:cb_convol_model}
    , 
\end{align}
avec $\hat{\mathcal{H}}_{f, f', t'} \triangleq \sum_{m=-N+1}^{N-1} \hat{h}(t'L - m) W_{f,f'}(m)$ et les notations de Eq.~\eqref{eq:cb_convol_model} coïncidant à celles de l'Eq.~(\ref{eq:full_convolution}-\ref{eq:big_window}).
Suivant~\cite{avargel_system_2007}, nous fixons le nombre de bandes de convolution inter-bande $F'$ à $4$.

Notre fonction de coût de correspondance de réverbération correspond à l'erreur moyenne quadratique pour le problème de déconvolution. 
Un terme de régularisation est ajouté pour encourager les log-amplitudes du signal reverbérant estimé à se rapprocher de celles de la vérité terrain, et la fonction de coût d'entraînement du modèle est, avec $\lambda=\gamma=1$ comme dans~\cite{schwarMultiScaleSpectralLoss2023}:

\begin{equation}
 \mathcal{L} = \sum_{f,t} \left[\lvert \hat{Y}_{f,t}  - Y_{f,t} \rvert^2 + \lambda \left| \log\left(\frac{1 + \gamma \lvert \hat{Y}_{f,t}\rvert}{1 + \gamma \lvert Y_{f,t} \rvert} \right) \right|^2 \right]
\end{equation}

\section{Expériences}

Nous comparons notre méthode de déréverbération non-supervisée avec celle utilisée par MetricGAN-U.

\subsection{Variants de RNP}

Nous évaluons plusieurs variantes de notre méthode avec FullSubNet (FSN)~\cite{hao_fullsubnet_2021}. 
Il a été déjà combiné avec des stratégies d'entraînement informées par la réverbération~\cite{zhou_2023_reverb_time_shortening}.
Nous considérons aussi le modèle baseline BiLSTM~\cite{weningerSpeechEnhancementLSTM2015} utilisé comme générateur dans MetricGAN-U.
Ce modèle est beaucoup plus simple puisqu'il permet de traiter seulement des masques d'amplitude et servira d'indicateur pour le comportement de notre méthode avec un modèle moins expressif.

\subsection{Variantes de supervision}

Nous considérons différentes variantes de supervision: 
\par\noindent \underline{Supervision Forte}:
Ce variant correspond à la fonction de coût originale de chacun des modèles, requérant des paires de signaux.
Le BiLSTM est entraîné en utilisant l'erreur moyenne quadratique entre les spectrogrammes d'amplitude secs et déréverbérés. FSN est entraîné à minimiser la distance euclidienne entre le masque complexe idéal et estimé (cRM).

\par\noindent \underline{Supervision faible} :
Ce variant d'auto-supervision par la réverbération correspond à notre modèle de RIS, dont le modèle d'analyse de paramètres acoustiques est un modèle oracle.
Suivant des expériences conduites dans~\cite{bahrmanHybridModelWeaklySupervised2025}, où il a été montré que cela n'impactait que peu la performance de déréverbération, nous fixons le temps de mixage $n_m$ et $\sigma$ à la valeur moyenne sur la base de données. 
Pour $n_m$ cela correspond au temps de mixage moyen de notre base de données selon la formule décrite dans~\cite{blesser2001synthesisOfReverbViewPoints}, soit $20$~ms, ou $n_m=0.02 f_s$.
Le paramètre sigma est fixé à 
$0.02$.
Ainsi pour ce variant seul le $\rtsixty$ est calculé de manière non aveugle à partir de la RIS.

\par\noindent \underline{Auto-supervision par la réverbération (aveugle)} :
Ce variant utilise l'algorithme d'estimation aveugle du $\rtsixty$ fondée sur la décomposition en sous-bandes du spectrogramme du signal réverbérant décrite dans~\cite{prego_2015_blind}.
L'algorithme est calibré sur 100 couples $(y,\rtsixty)$.

\par\noindent \underline{Auto-supervision par une métrique (SRMR)} :
Nous considérons aussi la baseline de MetricGAN-U correspondant au modèle BiLSTM supervisé par la métrique du SRMR.

\subsection{Configuration d'entrainement}

Comme pour FullSubNet original, des extraits de parole réverbérante de 49151 échantillons (environ 3 secondes à 16 kHz) sont traités dans le plan TFCT en utilisant une fenêtre de Hann de taille 512 avec un pas de $50~\%$. 
Nous utilisons l'optimiseur Adam et arrêtons l'entraînement selon l'évolution de la métrique SISDR sur un set de validation. 

\subsection{Données}
Comme pour \cite{bahrmanHybridModelWeaklySupervised2025},
nous avons simulé un ensemble de données d'entraînement en convoluant dynamiquement des signaux de parole sèche avec des RIS simulées. 
Les signaux de parole sèche sont échantillonnés de manière aléatoire à partir des enregistrements du microphone de casque de WSJ1~\cite{wsj1}. 
L'ensemble d'entraînement représente 73 heures cumulées d'audio divisés en 
60307 extraits. 
L'ensemble des RIS simulées se compose de 32 000 RIS tirées de 2 000 pièces simulées à l'aide de la méthode de source-image de pyroomacoustics~\cite{scheibler_pyroomacoustics_2018}.
Les dimensions de la pièce et le RT60 sont uniformément échantillonnés les intervalles de $[5,10] \times [5,10] \times [2,5,4]~\text{m}^3$, et $[0,2,1,0]$~s.
La distance source-microphone est uniformément distribuée dans $[0.75, 2.5]$~m, 
et la source et le microphone sont tous deux à au moins $50$~cm des murs.
Afin d'aligner la cible du signal sec et le trajet direct, les échantillons de RIS précédant le trajet direct sont éliminés et celle-ci est normalisée de sorte que le trajet direct soit d'amplitude 1.

\section{Résultats et discussion}

Nous évaluons la performance de nos méthodes (supervision faible et aveugle) sur des locuteurs de WSJ et des salles non vues à l'entraînement.
La performance est évaluée à l'aide des métriques Scale-Invariant Signal to Distortion Ratio (SISDR), Extended Short-Time Objective Intelligibility (ESTOI), Wide-Band Perceptual Evaluation of Speech Quality (WB-PESQ), et SRMR.
Les résultats sont présentés dans le tableau~\ref{tab:results_v2}.
La ligne dénotée \og Réverbérant\fg correspond aux signaux non traités.
Toutes les variantes proposées présentent une amélioration des métriques SISDR, ESTOI et WB-PESQ, donc parviennent à déréverbérer la parole avec succès.
La baseline (BiLSTM+SRMR) excelle en termes de SRMR, mais cette performance est au détriment des résultats de SISDR et STOI, qui sont dégradés par rapport à l'entrée réverbérante.
Cela confirme le principal désavantage de la déréverbération auto-supervisée par une métrique, dans le sens où elle tend à n'optimiser que la métrique cible. 
En effet, toutes nos méthodes proposées performent mieux que la baseline sur toutes les autres métriques que le SRMR.
Cela démontre la supériorité de l'auto-supervision par la réverbération sur l'auto-supervision par la métrique.
De plus, les performances des variantes de supervision aveugles, n'ayant pas accès au $\rtsixty$ oracle, sont très proches de la performance de la supervision faible.
Cela montre la robustesse de notre méthode à de faibles erreurs d'estimation de ce paramètre.
Enfin, en comparant les RNP, on remarque que les résultats du modèle BiLSTM sont moins dégradés par le passage de supervision forte à faible que ceux du modèle FSN.
Cela peut être expliqué par le fait que ce premier modèle est agnostique à la phase du signal réverbérant, particulièrement perturbée par notre modèle de réverbération.

\begin{table}[t]
  \centering
  \caption{Scores de déreverberation $\pm$ écart-type\\
  Pour chaque métrique, les meilleurs valeurs sont les plus hautes
  }
  \sisetup{
detect-weight, %
mode=text, %
tight-spacing=true,
round-mode=places,
round-precision=2,
table-format=1.2,
table-number-alignment=center
}
\resizebox{\linewidth}{!}{
\setlength{\tabcolsep}{3pt}
  \begin{tabular}{l|r|*{1}{S[round-precision=1,table-format=2.1]@{\,\( \pm \)\,}S[round-precision=1,table-format=1.1]S@{\,\( \pm \)\,}SS@{\,\( \pm \)\,}SS[round-precision=1,table-format=2.1]@{\,\( \pm \)\,}S[round-precision=1,table-format=1.1]S}}
  RNP  & \multicolumn{1}{c|}{Supervision} & \multicolumn{2}{c}{SISDR} & \multicolumn{2}{c}{ESTOI} & \multicolumn{2}{c}{WB-PESQ} & \multicolumn{2}{c}{SRMR}\\
  \hline
\multirow{3}{*}{FSN} & Forte & 5.622 & 3.858  & 0.838 & 0.097  & 2.545 & 0.669  & 8.228 & 3.488  \\
 & faible & 2.886 & 3.525  & 0.707 & 0.146  & 1.777 & 0.696  & 6.912 & 2.798  \\
 & Aveugle (Proposée) & 2.795 & 3.409  & 0.708 & 0.147  & 1.778 & 0.702  & 6.895 & 2.796  \\
 \hline
\multirow{3}{*}{BiLSTM} & Forte & 1.266 & 4.266  & 0.776 & 0.122  & 2.250 & 0.780  & 7.886 & 3.018  \\
 & faible & 1.557 & 3.672  & 0.708 & 0.149  & 1.837 & 0.742  & 6.865 & 2.802  \\
 & Aveugle (Proposée) & 1.534 & 3.674  & 0.708 &  0.149  & 1.836 & 0.742  & 6.866 & 2.801  \\
\hline
BiLSTM & SRMR (Baseline) & -1.509 & 3.455  & 0.638 & 0.180  & 1.776 & 0.720  & 10.887 & 4.341  \\
\hline
\multicolumn{2}{r|}{Réverbérant}   & -1.339 & 3.480  & 0.685 & 0.161  & 1.748 & 0.739  & 6.911 & 2.892  \\
       \end{tabular}
  }
  \label{tab:results_v2}
  \vspace{-1em}
\end{table}

\section{Conclusion}

Nous avons proposé une nouvelle approche non-supervisée pour la déréverbération de la parole, 
consistant à entraîner un réseau neuronal profond à prédire un signal sec à partir d'un signal réverbérant, de telle sorte qu'un modèle de réverbération appliqué sur cet estimé sec corresponde à l'entrée réverbérante. 
Cette méthode ouvre la voie vers une variété de techniques de déréverbération pour des scénarios où peu de données sont disponibles.
Les travaux futurs seront consacrés à l'application de ces travaux à des approches génératives auto-supervisées afin de mieux considérer le modèle de RIS probabiliste.

\subsubsection*{Remerciements}
 Ce travail a été financé par l’Union européenne (ERC, HI-Audio, 101052978). Les points de vue et les opinions exprimés sont ceux des auteurs et ne reflètent pas nécessairement ceux de l’Union européenne ou du Conseil européen de la recherche. Ni l’Union européenne ni l’organisme subventionnaire ne peuvent en être tenus pour responsables.

\bibliography{biblio}

\end{document}